\documentclass[letterpaper, 10 pt, conference]{ieeeconf}  

\IEEEoverridecommandlockouts                              

\makeatletter
\let\NAT@parse\undefined
\makeatother
\usepackage[breaklinks,citecolor=red,urlcolor=red,bookmarks=false,hypertexnames=true]{hyperref}
\usepackage{color}
\usepackage{xcolor}
\usepackage{graphicx}
\usepackage{caption}
\usepackage{subfigure}
\usepackage{makecell}
\usepackage{longtable}
\usepackage{multirow}
\usepackage{threeparttable}
\usepackage{booktabs}
\usepackage{mathtools}
\usepackage{tikz}
\usepackage[sort&compress,numbers]{natbib}
\usepackage{balance}

\usepackage{float}

\title{\LARGE \bf
Which Framework is Suitable for Online 3D Multi-Object Tracking for Autonomous Driving with Automotive 4D Imaging Radar?
}

\author{\authorblockA{
Jianan Liu$^{1*}$,
Guanhua Ding$^{2*}$,
Yuxuan Xia$^{3}$,
Jinping Sun$^{2}$, 
Tao Huang$^{4}$,
Lihua Xie$^{5}$, 
and Bing Zhu$^{6\dag}$
}
\thanks{This work has been submitted to the IEEE for possible publication. Copyright may be transferred without notice, after which this version may no longer be accessible.}
\thanks{$^{1}$Vitalent Consulting, Gothenburg, Sweden. Email: jianan.liu@vitalent.se.}
\thanks{$^{2}$The School of Electronics and Information Engineering, Beihang University, Beijing, P.R. China. Email: \{buaadgh, sunjinping\}@buaa.edu.cn
}
\thanks{$^{3}$The Department of Electrical Engineering, Linköping University, Linköping, Sweden. Email: yuxuan.xia@liu.se.}
\thanks{$^{4}$The College of Science and Engineering, James Cook University, Smithfield QLD 4878, Australia. Email: tao.huang1@jcu.edu.au.}
\thanks{$^{5}$The School of Electrical and Electronic Engineering, Nanyang Technological University, Singapore 639798. Email: elhxie@ntu.edu.sg.}
\thanks{$^{6}$The School of Automation Science and Electrical Engineering, Beihang University, Beijing, P.R. China. Email: zhubing@buaa.edu.cn.}
\thanks{$*$ Both authors contribute equally to the work and are co-first authors.}
\thanks{$\dag$ Corresponding author.
}
\thanks{This paper has been accepted by IEEE 35th Intelligent Vehicles Symposium (IV). Code is available at \url{https://github.com/dinggh0817/4D_Radar_MOT}.
}
}

\begin{document}

\maketitle

\begin{abstract}

Online 3D multi-object tracking (MOT) has recently received significant research interests due to the expanding demand of 3D perception in advanced driver assistance systems (ADAS) and autonomous driving (AD). Among the existing 3D MOT frameworks for ADAS and AD, conventional point object tracking (POT) framework using the tracking-by-detection (TBD) strategy has been well studied and accepted for LiDAR and 4D imaging radar point clouds. In contrast, extended object tracking (EOT), another important framework which accepts the joint-detection-and-tracking (JDT) strategy, has rarely been explored for online 3D MOT applications. This paper provides the first systematic investigation of the EOT framework for online 3D MOT in real-world ADAS and AD scenarios. Specifically, the widely accepted TBD-POT framework, the recently investigated JDT-EOT framework, and our proposed TBD-EOT framework are compared via extensive evaluations on two open source 4D imaging radar datasets: View-of-Delft and TJ4DRadSet. Experiment results demonstrate that the conventional TBD-POT framework remains preferable for online 3D MOT with high tracking performance and low computational complexity, while the proposed TBD-EOT framework has the potential to outperform it in certain situations. However, the results also show that the JDT-EOT framework encounters multiple problems and performs inadequately in evaluation scenarios. After analyzing the causes of these phenomena based on various evaluation metrics and visualizations, we provide possible guidelines to improve the performance of these MOT frameworks on real-world data. These provide the first benchmark and important insights for the future development of 4D imaging radar-based online 3D MOT algorithms. 

\end{abstract}

\section{Introduction}
\label{introduction}

Online 3D multi-object tracking (MOT) is a critical component in advanced driver assistance systems (ADAS) and autonomous driving (AD) applications. It helps the autonomous vehicle to achieve robust and accurate 3D perception by eliminating uncertainties in data association and multi-object state estimation. Due to the advances in sensor and signal processing technology, online 3D MOT using various types of sensors, e.g., camera, LiDAR, and radar, has received substantial interests in recent years \cite{Camera_MOT, PC3T, Radar_MOT}.

Among all commonly available sensor modalities, automotive radars, the only cost-effective sensor that can operate in both extreme lighting conditions and adverse weather \cite{Radar_perception_survey}, have been widely adopted for perception tasks including instance segmentation \cite{radar_instance_segmentation_1, radar_instance_segmentation_2}, object detection \cite{Radar_heatmap_object_detection_and_point_target_tracking, Radar_LiDAR_fusion_object_detection_RaLiBEV}, and MOT \cite{Radar_Tracking_ITSC_2022}. Although conventional automotive radars can effectively separate objects in range and Doppler velocity dimensions, the low angular resolution of radar measurements limits the performance of radar-based object detection and MOT. Recently, the 4D imaging radar based on the multiple-input multiple-output (MIMO) technology attracts increasing attention \cite{intro_4d_radar, 4D_radar_overview}. Unlike conventional automotive radars, 4D imaging radars are capable of measuring the range, velocity, azimuth, and elevation of an object, thereby providing new possibilities to develop novel radar-based 3D MOT methods.

The design paradigms of 3D MOT methods can be divided into two categories: model-based and deep learning-based \cite{model_based_and_DL_based_MOT, 3DMOTFormer}. The model-based paradigm employs meticulously designed multi-object dynamic and measurement models, making it suitable for the development of efficient and robust 3D MOT methods. 
As a typical framework of the model-based MOT paradigm, the point object tracking (POT) framework using tracking-by-detection (TBD) strategy has been widely adopted in academia and industry\cite{ACK3DMOT, SimpleTrack, PF-MOT, LiDAR_based_point_target_MOT_LEGO, LiDAR_based_point_target_MOT_PMBM, LiDAR_based_point_target_MOT_GNN-PMB}. POT assumes that each object generates at most one measurement per sensor scan; however, a 3D object often generates multiple measurement points in LiDAR and 4D imaging radar point clouds. Consequently, object detection is performed before tracking to combine the measurements generated by the same object into a single detection. The effectiveness of the TBD-POT framework has been validated in several real-world LiDAR-based online 3D MOT tasks \cite{PC3T, ACK3DMOT, SimpleTrack, PF-MOT, LiDAR_based_point_target_MOT_LEGO, LiDAR_based_point_target_MOT_PMBM, LiDAR_based_point_target_MOT_GNN-PMB}. 

Another model-based MOT framework receiving increasing attention in the tracking literature is extended object tracking (EOT)
\cite{Extended_Target_Tracking_Overview, Extended_Target_MOT_GGIW-CPHD, Extended_Target_MOT_GGIW-PMBM, Extended_Target_MOT_GGIW-PMB, Extended_Target_MOT_GGIW-PMBM_2, Extended_Target_MOT_EO-GM-PHD, Extended_Target_MOT_BGGIW}. In contrast to POT, EOT assumes that an object can generate multiple measurements per sensor scan. Therefore, EOT can achieve joint-detection-and-tracking (JDT) without an additional object detection module and is claimed to achieve promising results for single object tracking using real-world LiDAR point clouds \cite{LiDAR_Extended_Target_SOT_Rectangle_with_reference_points_GM-PHD_1,LiDAR_Extended_Target_SOT_Rectangle_with_reference_points_GM-PHD_2,LiDAR_3D_Extended_Target_SOT_1, LiDAR_3D_Extended_Target_SOT_2} and automotive radar detection points \cite{radar_3D_Extended_Target_SOT, radar_3D_Extended_Target_SOT_2}. However, EOT has rarely been conducted for online 3D MOT in complex ADAS and AD scenarios with real-world data. Currently, only two available works attempt to evaluate the EOT framework for real-world LiDAR-based MOT \cite{LiDAR_3D_Extended_Target_MOT_1, LiDAR_3D_Extended_Target_MOT_2}. None of the aforementioned works provided detailed performance of tracking multiple objects with different classes in an ADAS/AD dataset, nor did they perform a systematic analysis using widely accepted metrics. Thus, the applicability of EOT in complex ADAS and AD scenarios has not really been demonstrated. Moreover, with the rapid development of deep learning, almost all the state-of-the-art approaches for 3D MOT with point clouds in ADAS and AD scenarios follow either the TBD-POT framework using deep learning-based object detector \cite{PC3T, LiDAR_based_point_target_MOT_LEGO, ACK3DMOT, SimpleTrack, PF-MOT, LiDAR_based_point_target_MOT_PMBM, LiDAR_based_point_target_MOT_GNN-PMB}, or the deep learning-based tracking paradigm \cite{SimTrack, CenterTube, 3DMODT, MUTR3D, Trackformer}, which seems to imply that EOT is no longer necessary. Specifically, it remains an open question that whether EOT can outperform the traditional TBD-POT framework for 3D MOT with point clouds in terms of performance and complexity. In this study, this open question is answered for the first time with comprehensive evaluations and analyses. 


Specifically, the contributions of this paper are:
\begin{itemize}
\item This paper provides the first benchmark for subsequent studies on 4D imaging radar-based online 3D MOT in ADAS and AD by comparing POT and EOT frameworks. The evaluations reveal the pros and cons of POT and EOT frameworks, while our analyses provide guidelines for designing online 3D MOT algorithms.

\item To fill the gap between theory and practice for EOT-based online 3D MOT, for the first time the EOT framework is systematically investigated 
in real-world ADAS and AD scenarios. While the extensively investigated JDT-EOT framework performs inadequately, our proposed TBD-EOT framework, which leverages the strength of deep learning-based object detector, achieves superior tracking performance and computational efficiency compared with the JDT-EOT framework.

\item Experiment results indicate that the conventional TBD-POT framework remains preferable for online 3D MOT with 4D imaging radar due to its high tracking performance and computational efficiency. However, the TBD-EOT framework can outperform TBD-POT in certain situations, demonstrating the potentials of applying EOT for online 3D MOT in real-world ADAS and AD applications.

\end{itemize}

The rest of the paper is organized as follows. Related works are reviewed in Section \ref{relatedwork}. 
Three different online 3D MOT frameworks based on POT and EOT are explained in Section \ref{proposed methods}. 
Evaluation results of the investigated 3D MOT frameworks on View-of-Delft and TJ4DRadSet datasets are provided and compared systematically in Section \ref{result}. 
Finally, the conclusions are drawn in Section \ref{conclusion}.

\section{Related Works}\label{relatedwork}
\subsection{3D Object Detection with 4D Imaging Radar}
Due to the limited angular resolution and multi-path effect, the 4D imaging radar point cloud is sparser and contains more noise and ambiguities compared to LiDAR. To address these issues, several neural network-based 3D object detection methods for 4D imaging radar have recently been proposed. For example, a self-attention mechanism in RPFA-Net \cite{4d_imaging_radar_3d_object_detection_2} is employed to extract global features from 4D radar point clouds, achieving improved performance for estimating object heading angles. A 3D object detection framework is proposed in \cite{4d_imaging_radar_3d_object_detection_3} to accumulate temporal and spatial features in multiple 4D radar frames through velocity compensation and inter-frame matching. Multiple representations for 4D radar points are introduced in SMURF \cite{SMURF} by utilizing pillarization and kernel density estimation techniques, achieving state-of-the-art performance on two latest 4D imaging radar datasets, VoD \cite{VoD_dataset} and TJ4DRadSet \cite{TJ4DRadSet_dataset}. Moreover, 4D imaging radar are also fused with camera \cite{4d_imaging_radar_camera_3d_object_detection_rcfusion, 4d_imaging_radar_camera_3d_object_detection_LXL} and LiDAR \cite{4d_imaging_radar_lidar_3d_object_detection_1, 4d_imaging_radar_lidar_3d_object_detection_2} for performance improvement.


\subsection{3D Multi-Object Tracking with LiDAR}

The majority of LiDAR-based 3D MOT methods employ a traditional TBD strategy, in which an object detector processes the point cloud to produce detection results in the form of bounding boxes, and then a point object tracker performs MOT with the detection. As many 3D object detectors for LiDAR are sufficiently accurate, adequate tracking performance can be achieved by using a simple Bayesian MOT algorithm such as the global nearest neighbour tracker with heuristic track management \cite{PC3T, ACK3DMOT, SimpleTrack, PF-MOT}. However, due to the fact that the detectors still produce false detections, these MOT methods could suffer from track fragmentation and object ID switches. Several random finite set (RFS)-based methods have been proposed in an effort to further enhance the tracking performance. RFS-$M^3$ \cite{LiDAR_based_point_target_MOT_PMBM} employs a Poisson multi-Bernoulli mixture (PMBM) filter in conjunction with a neural network-based 3D object detector. Liu et al. further modify PMBM and propose Poisson multi-Bernoulli (PMB) filter with global nearest neighbour (GNN-PMB) \cite{LiDAR_based_point_target_MOT_GNN-PMB} as a simple and effective online MOT algorithm for LiDAR. A novel MOT framework based on the sum-product algorithm \cite{Point_target_MOT_BP} is proposed to achieve efficient probabilistic data association and substantially reduces ID switch errors. 

\begin{figure*}[tb]
\centering
\includegraphics[width=0.95\textwidth]{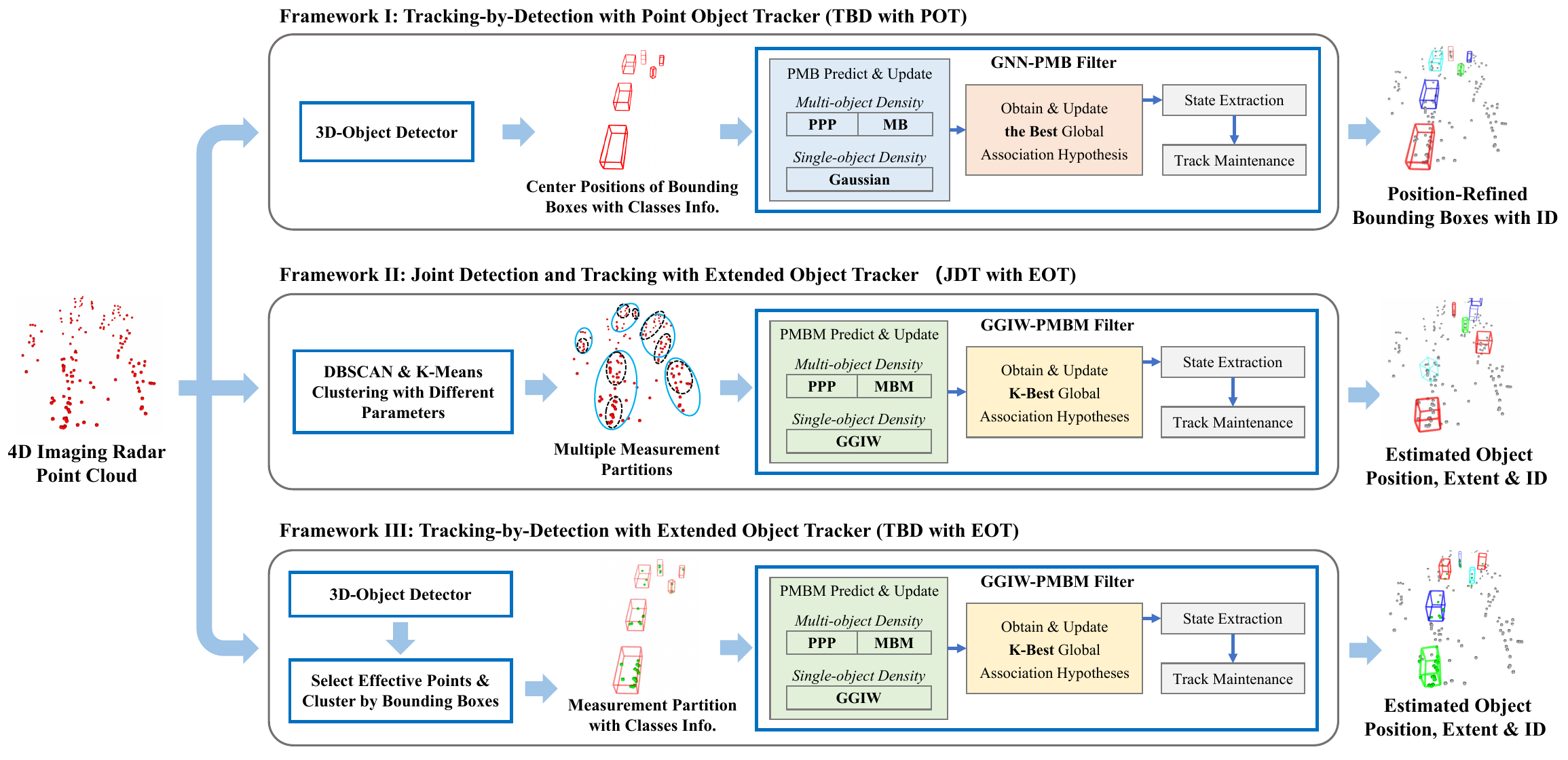}
\caption{The illustration of three different frameworks for online 3D MOT with 4D imaging radar point cloud.}
\label{Scheme1_2_3}
\end{figure*}

On the other hand, the LiDAR-based 3D MOT with JDT strategy has been implemented using the deep learning-based paradigm 
as well. For instance, SimTrack \cite{SimTrack} integrates data association and track management in an end-to-end trainable model
, CenterTube \cite{CenterTube} achieves JDT by detecting 4D spatio-temporal tubelets in point cloud sequences, 3DMODT \cite{3DMODT} can directly operate on raw LiDAR point clouds and employs an attention-based refinement module for affinity matrices. However, the EOT framework under the model-based paradigm, which can also accept the JDT strategy, has rarely been investigated for LiDAR-based 3D MOT in real-world ADAS and AD scenarios. Thus, it remains an area requiring further research. 

\section{Methodologies}      
\label{proposed methods}

In this section, we introduce three different frameworks for online 3D MOT with 4D imaging radar point clouds, including TBD-POT, JDT-EOT, and our proposed TBD-EOT, see Fig. \ref{Scheme1_2_3} for an illustration.

\subsection{Framework I: TBD with POT}

The TBD-POT framework has been widely adopted in literature for MOT with different sensor modalities, e.g. \cite{PC3T, PF-MOT, CAMO-MOT, C_R_Track_Level_Fusion}. In this tracking framework, the 4D imaging radar point cloud is first processed by an object detector to generate 3D bounding boxes that provide information such as object position, bounding box size, orientation, class, and detection score. The POT algorithm often takes two-dimensional object position measurements in Cartesian coordinate and performs MOT on the bird-eye's view (BEV) plane to simplify calculations. Other information of the 3D bounding boxes is then combined with estimated object positions and IDs to generate 3D tracking results. The TBD-POT framework has two main advantages: 1) the POT algorithm can leverage the extra information such as object classes and detection scores to further improve the tracking performance; 2) POT is typically less compute-intensive compared to EOT.

The GNN-PMB filter \cite{LiDAR_based_point_target_MOT_GNN-PMB}, one of the state-of-the-art POT approaches for LiDAR-based online 3D MOT, is selected as the POT algorithm. The filter estimates the multi-object state by propagating a PMB density over time, which combines a Poisson point process (PPP) for modeling undetected objects and a multi-Bernoulli (MB) process for modeling detected objects. The data association is achieved by managing local and global hypotheses. For each time step, a measurement can be matched with a previously tracked object, a newly detected object, or a false alarm to generate different local hypotheses with corresponding costs. Then, a group of compatible local hypotheses are collected in a global hypothesis \cite{global_hypothesis_definition}, which defines a possible association between existing objects and measurements. 
Finally, the optimal data association result, which is the global hypothesis with the lowest total cost, is obtained by solving the 2D-assignment problem on the cost matrix. Different from the PMBM filter that calculates and propagates multiple global hypotheses, GNN-PMB only propagates the best global hypothesis to reduce computational complexity without significantly deteriorating the tracking performance \cite{LiDAR_based_point_target_MOT_GNN-PMB}. In summary, the first online 3D MOT framework in this paper combines a deep learning-based 3D object detector with the GNN-PMB filter, as illustrated in the first row of Fig. \ref{Scheme1_2_3}.

\subsection{Framework II: JDT with EOT}

In contrast to the first framework, the JDT-EOT framework operates on 4D radar point clouds by detecting and tracking multiple objects simultaneously. The point clouds go through a gating and clustering process to generate the measurement partition (a group of disjoint clusters); then, an EOT filtering algorithm performs 3D MOT 
using these clusters. To reduce computational complexity, clusters can be matched with objects during the gating process only if they fall within a distance threshold around the predicted position of the object. Clusters that cannot match with any existing object are then assigned to newborn objects or considered as clutter. Theoretically, this framework has the potential to provide more accurate estimates of the object position and shape while also reducing false negatives, since the point clouds contain more information than pre-processed 3D bounding boxes. However, it is challenging to produce proper measurement partitions, particularly for 4D radar point clouds with lots of ambiguities and clutters. As the distribution and density of point clouds can vary between objects, different clustering algorithms, such as DBSCAN \cite{DBSCAN} and $k$-means \cite{k-means}, with different parameter settings are usually employed to generate as many different measurement partitions as possible. This further increases EOT's computational complexity and poses a challenge to the real-time performance of this framework. 

In order to implement the JDT-EOT framework, as illustrated in the second row of Fig. \ref{Scheme1_2_3}, we select the PMBM filter with the gamma Gaussian inverse Wishart model (GGIW-PMBM), which is recognized as one of the state-of-the-art EOT algorithms due to its high estimation accuracy and manageable computational complexity \cite{Extended_Target_MOT_GGIW-PMBM, Extended_Target_MOT_GGIW-PMBM_2}. The PMBM filter \cite{point_target_MOT_PMBM} models object-originated measurements with the multi-Bernoulli mixture (MBM) density and propagating multiple global hypotheses to contend with the uncertainty of data association. The GGIW model assumes that the number of object-generated measurements is Poisson distributed, and the single measurement likelihood is Gaussian. Under this 
assumption, 
each object has an elliptical shape represented by the inverse Wishart (IW) density, 
while the major and minor axes of this ellipse can be used to form a rectangular bounding box. This 
simple but flexible extent modeling is sufficient to model different classes of objects \cite{Extended_Target_MOT_GGIW-CPHD, Extended_Target_MOT_GGIW-PMB, Extended_Target_MOT_GGIW-PMBM, Extended_Target_MOT_GGIW-PMBM_2}. More importantly, the GGIW implementation has the 
lowest computational complexity among all existing EOT implementations \cite{Extended_Target_Tracking_Overview}, which is desirable for real-time 3D MOT. Without detection bounding boxes and class information, the extent of a newborn object is initialized based on the spatial size of the associating cluster. Additionally, the extent estimates are processed by non-maximum suppression (NMS) \cite{NMS} to reduce the physically impossible overlapping tracking results.


\subsection{The Proposed Framework III: TBD with EOT}

To leverage the strengths of deep learning-based object detector and EOT, we present TBD-EOT as the third MOT framework. Instead of directly performing EOT on the radar point cloud, the points within detected bounding boxes are selected for clustering since these ``effective" points are more likely originated from objects than clutters. Compared with JDT-EOT, the advantage of TBD-EOT framework is twofold. First, the computational complexity of the data association in EOT can be substantially reduced by removing the clutter points, which leads to improved tracking performance with fewer false tracks. Second, the EOT algorithm can utilize the information from the detector to further improve the tracking performance, for example, by setting optimized parameters for different object classes and discarding bounding boxes with low detection scores. Compared with TBD-POT, the TBD-EOT framework employs a more realistic measurement model and has the potential to produce accurate object bounding boxes from extent estimates. As shown in the third row of Fig. \ref{Scheme1_2_3}, this MOT framework is implemented using the same 
3D object detector as the TBD-POT framework along with the GGIW-PMBM filter.

\section{Experiments and Analysis}
\label{result}

\subsection{Dataset and Evaluation Metrics}
\label{metrics}

We evaluate each online 3D MOT framework on two recently released 4D imaging radar-based autonomous driving datasets: View-of-Delft (VoD) \cite{VoD_dataset} and TJ4DRadSet \cite{TJ4DRadSet_dataset}. Both datasets contain synchronized 4D imaging radar, LiDAR, and camera data with high-quality annotations
. Each framework is evaluated with three object classes (car, pedestrian, and cyclist) on the validation set of VoD (sequence numbers 0, 8, 12, and 18) and part of the test set of TJ4DRadSet (sequence numbers 0, 10, 23, 31, and 41). These selected sequences cover various driving conditions and contain different classes of objects in balanced quantities. SMURF \cite{SMURF}, a state-of-the-art object detector for 4D imaging radar point clouds, is selected to extract bounding box detections for implementing TBD-POT and TBD-EOT. Since object class information is inaccessible for JDT-EOT, a heuristic classification step is employed in the state extraction procedure of this framework. In this step, unclassified tracking results are separated into cars, pedestrians, cyclists, and other objects based on the width and length of the estimated bounding boxes.

In the following evaluations, a set of commonly accepted MOT metrics for ADAS and AD are evaluated on the BEV plane, including multiple object tracking accuracy (MOTA), multiple object tracking precision (MOTP), true positive (TP), false negative (FN), false positive (FP), and ID switch (IDS). In addition, we employ a recently proposed MOT metric, higher order tracking accuracy (HOTA) \cite{HOTA}. HOTA decomposes into a family of sub-metrics, including detection accuracy (DetA), association accuracy (AssA), and localization accuracy (LocA), thus enabling a clear analysis of the MOT performance. 

\begin{table*}[ht]
\vspace{5pt}
\footnotesize
\caption{4D imaging radar-based 3D MOT tracking results on VoD validation set
 }
\label{comparison_with_different_trackers_3D_metric_vod_data}
\renewcommand\tabcolsep{4pt}
\begin{center}
\begin{threeparttable}
\begin{tabular}{|l|l|l|cccccccccc|}
  \hline
  Method & Framework & Class & HOTA$^\dag\uparrow$ & DetA$^\dag\uparrow$ & AssA$^\dag\uparrow$ & LocA$^\dag\uparrow$ & MOTA$^\dag\uparrow$ & MOTP$^\dag\uparrow$ & TP$\uparrow$ & FN$\downarrow$ & FP$\downarrow$ & IDS$\downarrow$\\
\hline
\multirow{3}{*}{\makecell[l]{SMURF +\\GNN-PMB}} & \multirow{3}{*}{\makecell[l]{TBD-POT}} & car & \textbf{54.36} & \textbf{42.64} & \textbf{69.34} & \textbf{93.90} & 36.26 & \textbf{93.60} & \textbf{2190} & \textbf{2101} & 593 & 41\\
& & pedestrian & 53.23 & \textbf{45.29} & 62.62 & \textbf{94.91} & \textbf{40.65} & \textbf{94.42} & \textbf{1925} & \textbf{1824} & \textbf{352} & 49\\
& & cyclist & 65.77 & \textbf{60.71} & 71.27 & \textbf{93.78} & \textbf{57.95} & \textbf{93.58} & 1045 & 389 & \textbf{201} & 13\\ \hline
\multirow{3}{*}{GGIW-PMBM} & \multirow{3}{*}{JDT-EOT} & car & 8.35 & 5.60 & 12.62 & 70.28 & -78.75* & 66.58 & 567 & 3724 & 3839 & 107\\
& & pedestrian & 16.21 & 7.40 & 36.00 & 89.28 & -9.79* & 90.92 & 326 & 3423 & 660 & 33\\
& & cyclist & 21.21 & 10.09 & 44.67 & 90.73 & -114.30* & 91.34 & 361 & 1073 & 1990 & 10\\ \hline
\multirow{3}{*}{\makecell[l]{SMURF +\\GGIW-PMBM}} & \multirow{3}{*}{\makecell[l]{TBD-EOT}} & car & 47.15 & 35.70 & 62.45 & 82.70 & \textbf{38.22} & 79.68 & 2145 & 2146 & \textbf{491} & \textbf{12}\\
& & pedestrian & \textbf{55.27} & 44.22 & \textbf{69.13} & 94.15 & 39.96 & 93.52 & 1906 & 1843 & 378 & \textbf{26}\\
& & cyclist & \textbf{66.47} & 58.48 & \textbf{75.64} & 92.68 & 54.32 & 92.25 & \textbf{1089} & \textbf{345} & 302 & \textbf{8}\\
\hline
\end{tabular}
\begin{tablenotes}
        \footnotesize
        \item[* The MOTA of GGIW-PMBM are negative values because there are significantly more FNs and FPs than TPs, while $\text{MOTA=1-(FN+FP+IDS)/(TP+FN)}$.]
        \item[$\dag$ The metrics are multiplied by 100. The bold values indicate the best results of each object class.]
\end{tablenotes}
\end{threeparttable}
\end{center}
\vspace{-5pt}
\end{table*}

\begin{table*}[ht]
\footnotesize
\caption{4D imaging radar-based 3D MOT tracking results on TJRadSet test set
 }
\label{comparison_with_different_trackers_3D_metric_tjradset_data}
\renewcommand\tabcolsep{4pt}
\begin{center}
\begin{threeparttable}
\begin{tabular}{|l|l|l|cccccccccc|}
  \hline
  Method & Framework & Class & HOTA$^\dag\uparrow$ & DetA$^\dag\uparrow$ & AssA$^\dag\uparrow$ & LocA$^\dag\uparrow$ & MOTA$^\dag\uparrow$ & MOTP$^\dag\uparrow$ & TP$\uparrow$ & FN$\downarrow$ & FP$\downarrow$ & IDS$\downarrow$\\
\hline
\multirow{3}{*}{\makecell[l]{SMURF +\\GNN-PMB}} & \multirow{3}{*}{\makecell[l]{TBD-POT}} & car & \textbf{43.41} & \textbf{32.19} & \textbf{59.32} & \textbf{89.50} & \textbf{24.56} & \textbf{88.36} & 961 & 1331 & \textbf{378} & 20\\
& & pedestrian & 31.21 & \textbf{27.96} & 34.82 & \textbf{96.00} & 20.17 & \textbf{95.95} & \textbf{294} & \textbf{638} & 99 & 8\\
& & cyclist & 42.22 & 35.43 & 50.32 & \textbf{93.28} & \textbf{23.74} & \textbf{92.82} & 448 & 542 & \textbf{200} & 13\\ \hline
\multirow{3}{*}{GGIW-PMBM} & \multirow{3}{*}{JDT-EOT} & car & 16.45 & 6.86 & 39.60 & 78.54 & -89.88* & 72.32 & 424 & 1868 & 2435 & 49\\
& & pedestrian & 13.19 & 10.03 & 17.41 & 93.52 & -132.30* & 94.52 & 239 & 693 & 1462 & 10\\
& & cyclist & 23.28 & 8.65 & 62.79 & 90.53 & -92.32* & 89.94 & 195 & 795 & 1105 & 4\\ \hline
\multirow{3}{*}{\makecell[l]{SMURF +\\GGIW-PMBM}} & \multirow{3}{*}{\makecell[l]{TBD-EOT}} & car & 38.16 & 28.35 & 51.88 & 82.19 & 24.39 & 78.55 & \textbf{962} & \textbf{1330} & 397 & \textbf{6}\\
& & pedestrian & \textbf{41.60} & 27.10 & \textbf{63.87} & 95.36 & \textbf{21.89} & 95.22 & 279 & 653 & \textbf{68} & \textbf{7}\\
& & cyclist & \textbf{49.48} & \textbf{36.42} & \textbf{67.23} & 92.05 & 20.10 & 91.46 & \textbf{505} & \textbf{485} & 298 & \textbf{8}\\
\hline
\end{tabular}
\begin{tablenotes}
        \footnotesize
        \item[* The MOTA of GGIW-PMBM are negative values because there are significantly more FNs and FPs than TPs, while MOTA=1-(FN+FP+IDS)/(TP+FN).]
        \item[$\dag$ The metrics are multiplied by 100. The bold values indicate the best result of each object class under each metric.]
\end{tablenotes}
\end{threeparttable}
\end{center}
\vspace{-5pt}
\end{table*}


Most notably, the MOTA, MOTP and HOTA metrics are calculated based on TP, FN, and FP, which are determined by the similarity score $S$ defined as:
\begin{equation}
    S=\max\left[0,\ 1-\frac{d(p,q)}{d_0} \right] \label{1}
\end{equation}
where $d(p,q)$ is the Euclidean distance between a object's estimated position $p$ and its corresponding ground-truth position $q$, and $d_0$ is the distance where $S$ reduces to zero. The pairs of estimation and ground-truth satisfying $S\geq\alpha$ are matched and counted as TPs, where $\alpha$ is the localization threshold. The remaining unmatched estimations become FPs, and unmatched ground-truths become FNs. The zero-distance is set to $d_0=4$m in following evaluations. For MOTA and MOTP, the localization threshold is set to $\alpha=0.5$. This setting indicates that an estimation can match with a ground-truth if the Euclidean distance between their center positions is no more than $2$m, which is aligned with the nuScenes \cite{nuScenes} tracking challenge\footnote{\href {https://www.nuscenes.org/tracking}{https://www.nuscenes.org/tracking}}. HOTA is calculated by averaging the results over different $\alpha$ values ($0.05$ to $0.95$ with an interval of $0.05$, as suggested in \cite{HOTA}).

\subsection{Comparison between Different Tracking Frameworks}
\label{sotacomp}
The evaluation results of the three online 3D MOT frameworks on VoD and TJ4DRadSet are provided in Table \ref{comparison_with_different_trackers_3D_metric_vod_data} and Table \ref{comparison_with_different_trackers_3D_metric_tjradset_data}, respectively. The hyper-parameters of the implemented algorithms, specifically, SMURF + GNN-PMB, GGIW-PMBM, and SMURF + GGIW-PMBM, are fine-tuned on the training sets by optimizing the HOTA metric.

\begin{figure}[h]
\footnotesize
\centering
\includegraphics[width=0.485\textwidth]{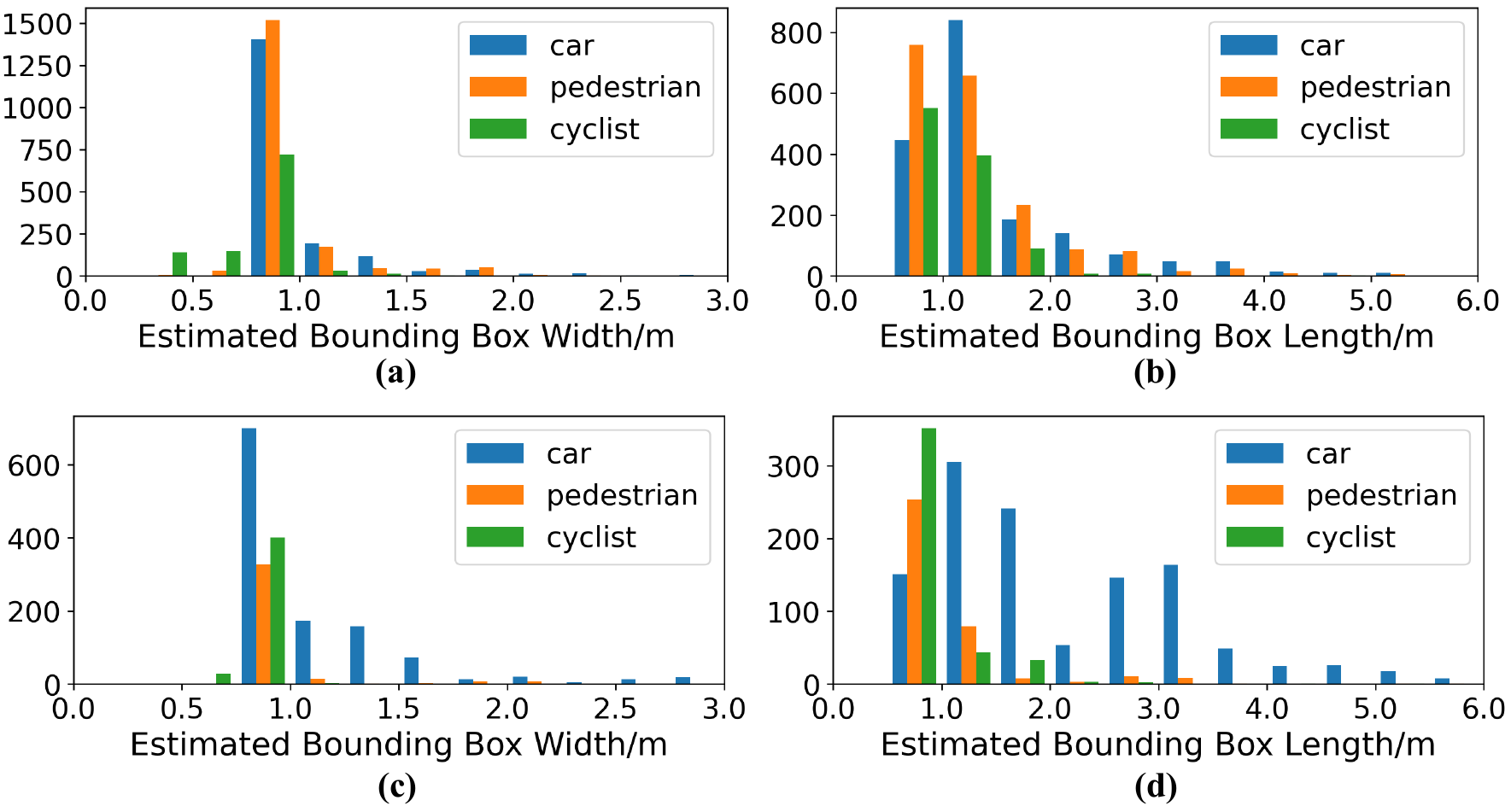}
\caption{Histogram of the TP bounding box size estimated by GGIW-PMBM. All tracking results within 2m of the ground truth positions are matched as TPs. (a) and (b) illustrate the width and length estimated on VoD validation set. (c) and (d) illustrate the width and length estimated on TJ4DRadSet test set.}
\label{histogram_of_bbox_size}
\vspace{-10pt}
\end{figure}

\emph{1) Performance of GGIW-PMBM:} 
Table \ref{comparison_with_different_trackers_3D_metric_vod_data} and Table \ref{comparison_with_different_trackers_3D_metric_tjradset_data} illustrate that the performance of GGIW-PMBM is
undesirable in our experiments. It is observed that GGIW-PMBM suffers from low detection accuracy for all three classes since the tracking results include significantly more FPs and FNs than TPs. To analyze the underlying reason, we re-calculate TP and FN using unclassified GGIW-PMBM tracking results, where any tracking results within $2$m of the ground-truth positions are matched as TPs. As shown in Table \ref{TP_FN_GGIW-PMBM}, the TPs for all three classes increase by large margins compared to the original evaluation results, demonstrating that GGIW-PMBM can produce tracking results close to the ground-truth positions. However, as illustrated in Fig. \ref{histogram_of_bbox_size}, a considerable portion of the TP bounding boxes estimated by GGIW-PMBM have similar width and length. Consequently, the heuristic classification step fails to classify some tracking results based on the estimated bounding box size, resulting in low detection accuracy in the original evaluation. 
\begin{table}[ht]
\footnotesize
\caption{TP and FN evaluated using unclassified GGIW-PMBM tracking results.
}
\renewcommand\arraystretch{1.2}
\label{TP_FN_GGIW-PMBM}
\renewcommand\tabcolsep{4pt}
\begin{center}
\begin{threeparttable}
\begin{tabular}{|l|l|p{35pt}<{\centering}|p{35pt}<{\centering}|p{55pt}<{\centering}|}
  \hline
  Dataset & Class & TP & FN & TP Increase (\%)\\
\hline
\multirow{3}{*}{VoD} & car & 1536 & 2755 & 170.90 \\
& pedestrian & 1703 & 2046 & 422.39\\
& cyclist & 988 & 446 & 173.68\\ \hline
\multirow{3}{*}{TJ4DRadSet} & car & 1157 & 1135 & 172.88\\
& pedestrian & 357 & 575 & 49.37\\
& cyclist & 430 & 560 & 120.51\\
\hline
\end{tabular}
\end{threeparttable}
\end{center}
\vspace{-5pt}
\end{table}

\begin{figure}[ht]
\centering
\includegraphics[width=0.38\textwidth]{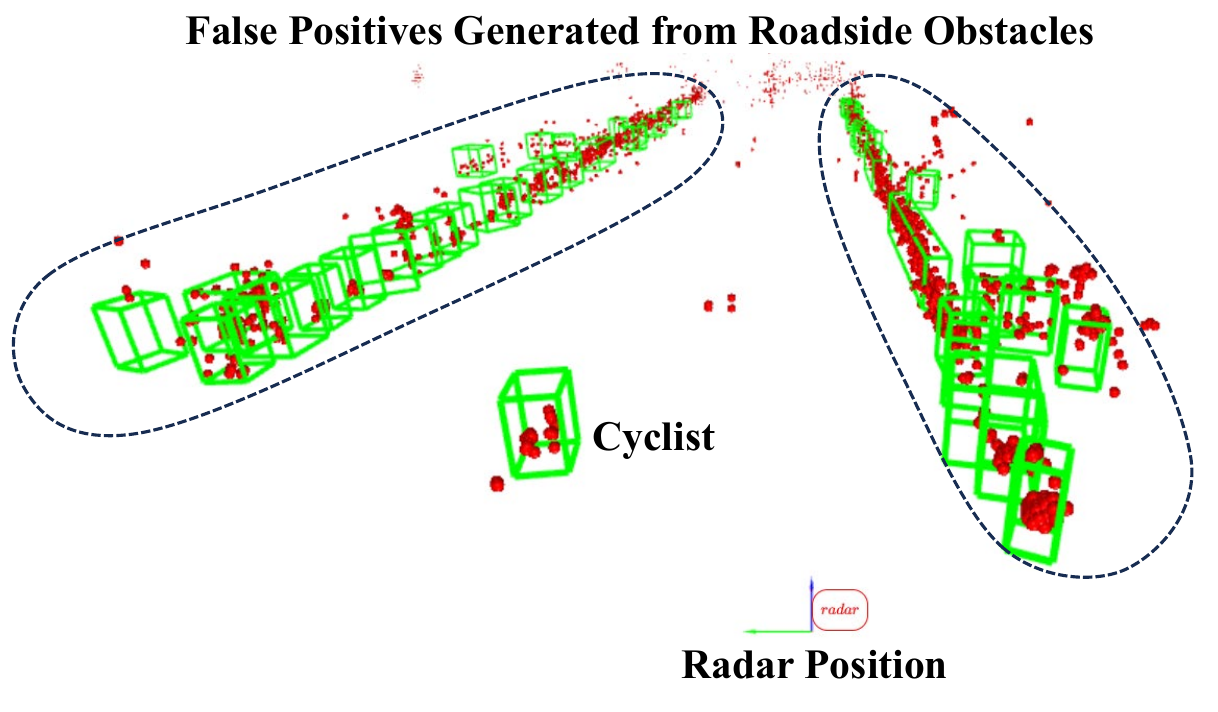}
\caption{The false positives generated by GGIW-PMBM from roadside obstacles in a scene of TJ4DRadSet test set. The red dots are radar points. The green boxes are estimated object bounding boxes.}
\label{roadside_clutters}
\end{figure}

We proceed to discuss how the performance of GGIW-PMBM differs between the two datasets. The MOTA for pedestrian class is substantially lower on TJ4DRadSet than on VoD, indicating that GGIW-PMBM generates more false tracks with small extent estimates on TJ4DRadSet. The disparity in performance can be attributed to the fact that the tested sequences of TJ4DRadSet contain dense clutters originating from roadside obstacles, whereas the clustering procedure is incapable of excluding these clutters. This effect is illustrated in Fig. \ref{roadside_clutters}, which displays a scene from TJ4DRadSet where the vehicle is travelling on a four-lane road with obstacles such as fences and street lights on both sides of the road. Since the roadside obstacles are stationary, this problem could be mitigated by removing radar points with low radial velocity prior to clustering. Supplementary experiments are not conducted here as TJ4DRadSet has not yet provided ego-vehicle motion data. However, such removal process can also influence the point clouds of stationary objects, thereby increasing the probability of these objects being mistracked.

In general, it can be inferred that GGIW-PMBM does not achieve superior performance than SMURF + GNN-PMB when applied to real-world 4D imaging radar point clouds. This is primarily due to the absence of object detector's information under the JDT-EOT framework, which makes it challenging to classify tracking results using heuristic methods and to distinguish object-generated point clouds from background clutters. 

\begin{figure}[t]
\centering
\includegraphics[width=0.45\textwidth]{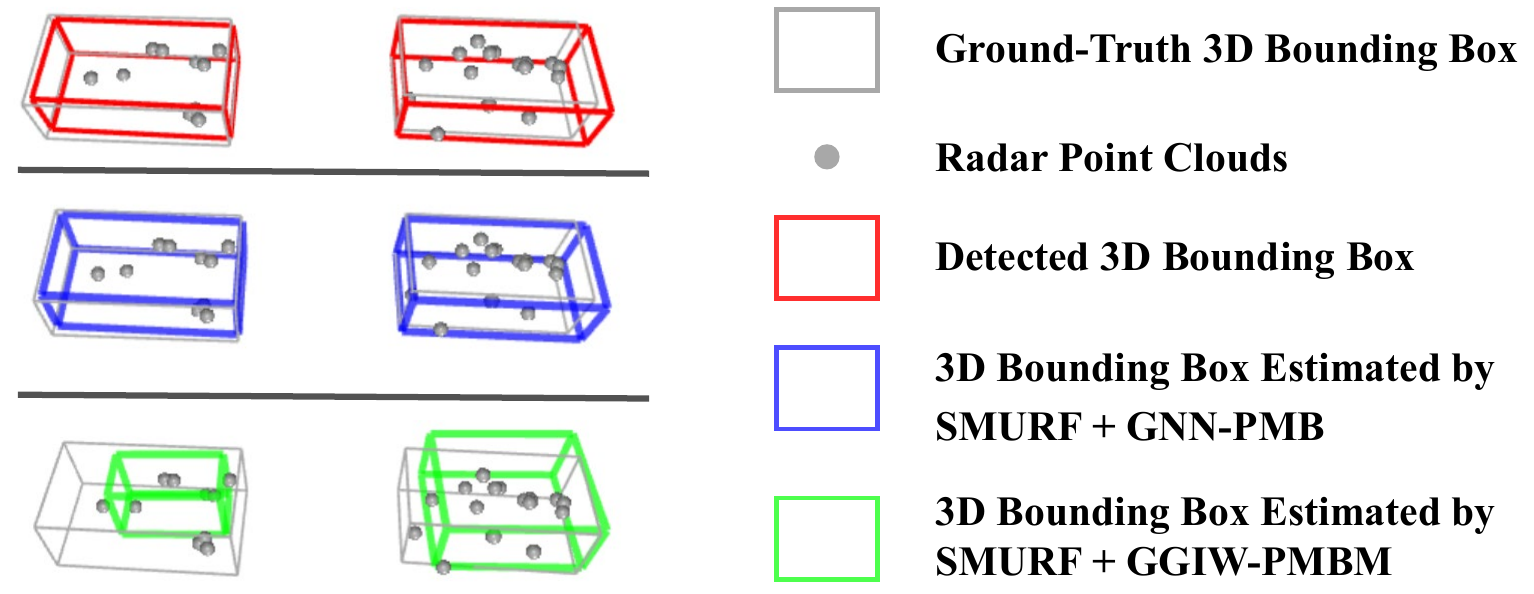}
\caption{The visualization of unevenly distributed radar point clouds for car objects taken from a scene in the VoD validation set. The figures in the left column illustrate the ground-truth, detected, and estimated 3D bounding boxes of two car objects at the same time step. The 4D radar is on the right side of the objects.}
\label{radar_point_distribution}
\vspace{-10pt}
\end{figure}

\emph{2) Performance of SMURF + GNN-PMB and SMRUF + GGIW-PMBM:} 
Different from GGIW-PMBM, both SMURF + GNN-PMB and SMURF + GGIW-PMBM utilize information from the object detector. As shown in Table \ref{comparison_with_different_trackers_3D_metric_vod_data} and Table \ref{comparison_with_different_trackers_3D_metric_tjradset_data}, SMURF + GNN-PMB outperforms SMURF + GGIW-PMBM in HOTA by a large margin with regard to car objects, primarily because the localization and detection accuracy of cars are notably lower for SMURF + GGIW-PMBM. To better illustrate this phenomenon, we evaluate the MOTA of car class under different localization thresholds $\alpha$, as shown in Table \ref{MOTA_with_different_alpha}. As $\alpha$ increases, the MOTA of SMURF + GGIW-PMBM decreases more rapidly than that of SMURF + GNN-PMB, indicating that more tracking results from SMURF + GGIW-PMBM are evaluated as FPs under the same TP matching criterion. The localization error of SMURF + GGIW-PMBM primarily stems from inaccuracies in modeling the distribution of point clouds. As shown in Fig. \ref{radar_point_distribution}, radar point clouds often congregate on the side of the car object nearest to the radar. This contrasts with the modeling assumption in the GGIW implementation, which assumes that the measurement points are distributed over the entire extent ellipse. Consequently, this discrepancy causes the estimated size and position of car objects to deviate from the ground-truth. Therefore, employing more accurate measurement models, e.g., Gaussian process \cite{LiDAR_3D_Extended_Target_SOT_2} and data-region association \cite{radar_3D_Extended_Target_SOT_2, PMRA-PMBM}, may improve the performance of TBD-EOT framework for large objects like cars. However, this could also increase the computational complexity.

\begin{table}[t]
\footnotesize
\caption{MOTA of car class evaluated under different localization thresholds $\alpha$.
}
\renewcommand\arraystretch{1.2}
\label{MOTA_with_different_alpha}
\renewcommand\tabcolsep{4pt}
\begin{center}
\begin{threeparttable}
\begin{tabular}{|p{20pt}<{\centering}|p{45pt}<{\centering}|p{45pt}<{\centering}|p{45pt}<{\centering}|p{45pt}<{\centering}|}
  \hline
  \multirow{2}{*}{$\alpha$} & \multicolumn{2}{c|}{SMURF + GNN-PMB}  & \multicolumn{2}{c|}{SMURF + GGIW-PMBM} \\ \cline{2-5}
  & VoD & TJ4DRadSet & VoD & TJ4DRadSet \\
\hline
0.5 & 36.26 & 24.56 & 38.22 & 24.39\\
0.6 & 36.03 & 23.95 & 34.68 & 17.67\\
0.7 & 35.40 & 21.86 & 22.12 & 5.19\\
0.8 & 34.16 & 12.00 & -8.72 & -15.27\\
\hline
\end{tabular}
\end{threeparttable}
\end{center}
\end{table}

\begin{figure}[ht]
\centering
\includegraphics[width=0.483\textwidth]{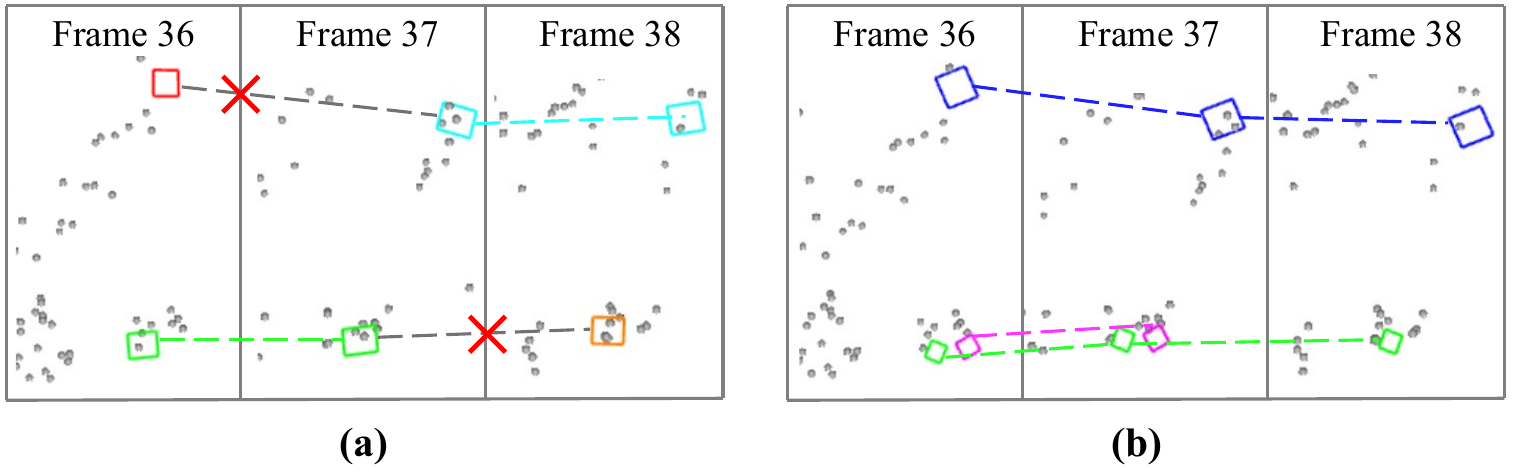}
\caption{Track ID maintenance for pedestrians in a scene of the VoD validation set. The tracking results of SMURF + GNN-PMB and SMURF + GGIW-PMBM are shown in (a) and (b), respectively. The dashed lines connect bounding boxes of the same object and the cross marks represent ID switches.}
\label{IDS}
\end{figure}

In terms of pedestrians and cyclists, it is noteworthy that SMURF + GGIW-PMBM outperforms SMURF + GNN-PMB on HOTA mainly due to its superior association accuracy (AssA). In addition, SMURF + GGIW-PMBM produces fewer IDS than SMURF + GNN-PMB for pedestrian and cyclist classes, as illustrated in Fig. \ref{IDS}. These phenomena are analyzed as below. 
First, GGIW-PMBM employs an adaptive detection model for the object \cite[Eq. (35)]{Extended_Target_MOT_GGIW-PMBM}. The object detection probability $P_d$ can be factorized as $P_d=P_{dm}P_{m}$, where $P_{m}$ denotes the probability of 
an existing object being measurable, i.e., the object generates a bounding box for GNN-PMB or at least one radar point for GGIW-PMBM; $P_{dm}$ represents the detection probability of a measurable object. In contrast to GNN-PMB which models $P_{m}$ as a fixed hyper-parameter, GGIW-PMBM calculates $P_{m}$ based on GGIW densities and more reliably estimates the object detection probability. 
Second, beside object position, the number and spatial distribution of object-originated radar points are also employed in the GGIW-PMBM filter to calculate the likelihood of association hypothesis. Since the GGIW density can model the distribution of radar points more accurately for small objects (as the points are less likely to congregate on one side of these objects compared to cars), GGIW-PMBM can utilize more 
information from point clouds of pedestrians and cyclists 
to accurately estimate $P_d$ and association hypothesis likelihood. Third, GGIW-PMBM propagates multiple global hypotheses over time, which potentially improves the data association accuracy with noisy radar measurements. These factors could help SMURF + GGIW-PMBM to achieve superior performance on AssA and IDS by reducing associations with false alarms and the false-termination of tracks.

Finally, the computational complexity for the three MOT algorithms is evaluated by the average frames processed per second (FPS). As shown in Table \ref{avg_run_time}, the average FPS of SMURF + GGIW-PMBM is about 20\% that of SMURF + GNN-PMB, while GGIW-PMBM is substantially slower than the other two algorithms, primarily due to the excessive number of possible measurement partitions generated from raw 4D imaging radar point clouds.

\begin{table}[t]
\footnotesize
\caption{FPS for MOT frameworks, evaluated from Python implementations with AMD 7950X CPU and 64 GB RAM.
}
\renewcommand\arraystretch{1.2}
\label{avg_run_time}
\renewcommand\tabcolsep{4pt}
\begin{center}
\begin{threeparttable}
\begin{tabular}{|l|c|}
  \hline
  Method & Avg. FPS in VoD / TJ4DRadSet\\
\hline
SMURF + GNN-PMB & 401.97 / 326.57 \\
GGIW-PMBM & 3.58 / 1.24 \\
SMURF + GGIW-PMBM & 81.59 / 87.95 \\ \hline
\end{tabular}
\end{threeparttable}
\end{center}
\vspace{-10pt}
\end{table}

\section{Conclusion and Future Work}
\label{conclusion}

This paper systematically compares the POT and EOT frameworks for online 3D MOT with 4D imaging radar point clouds on the VoD and TJ4DRadSet datasets. Three MOT frameworks, including TBD-POT, JDT-EOT, and TBD-EOT, are implemented with state-of-the-art methods and evaluated for car, cyclist, and pedestrian objects with commonly accepted 3D MOT metrics. Experiment results show that the traditional TBD-POT framework remains effective because its implementation, SMURF + GNN-PMB, achieves the best tracking performance for cars with the lowest computational complexity. However, the GGIW-PMBM implementation of the intensively studied JDT-EOT framework does not yield satisfactory tracking performance, primarily due to the incapability of conventional clustering methods 
to remove dense clutters and the high computational complexity caused by an excessive number of measurement partition hypotheses. Under the proposed TBD-EOT framework, SMURF + GGIW-PMBM shows great potential to outperform the implementation of TBD-POT, SMURF + GNN-PMB, by achieving superior association accuracy and more reliable ID estimation for both pedestrian and cyclist objects. Yet, the performance of SMURF + GGIW-PMBM deteriorates for cars because GGIW is unsuitable for modeling unevenly distributed radar point clouds, indicating the necessity of developing more realistic, accurate, and computationally efficient object models in the future.








\footnotesize



\begin{thebibliography}{60}
\bibitem{Camera_MOT} 
P. Li, and J. Jin, ``Time3D: End-to-end joint monocular 3D object detection and tracking for autonomous driving," in \emph{Proceedings of the IEEE/CVF Conference on Computer Vision and Pattern Recognition (CVPR)}, 2022, pp. 3885-3894.

\bibitem{Radar_MOT}
K. Shi, Z. Shi, C. Yang, S. He, J. Chen, and A. Chen, ``Road-map aided GM-PHD filter for multivehicle tracking with automotive radar," \emph{IEEE Transactions on Industrial Informatics}, vol. 18, no. 1, pp. 97-108, Jan. 2022.

\bibitem{PC3T} 
H. Wu, W. Han, C. Wen, X. Li, and C. Wang, ``3D multi-object tracking in point clouds based on prediction confidence-guided data association,'' \emph{IEEE Transactions on Intelligent Transportation Systems}, vol. 23, no. 6, pp. 5668-5677, Jun. 2022.

\bibitem{Radar_perception_survey} 
A. Pandharipande et al., ``Sensing and machine learning for automotive perception: A review," \emph{IEEE Sensors Journal}, vol. 11, no. 23, pp. 11097-11115, Jun. 2023.

\bibitem{radar_instance_segmentation_1}
J. Liu et al., ``Deep instance segmentation with automotive radar detection points,'' \emph{IEEE Transactions on Intelligent Vehicles}, vol. 8, no. 1, pp. 84-94, Jan. 2023.

\bibitem{radar_instance_segmentation_2}
W. Xiong, J. Liu, Y. Xia, T. Huang, B. Zhu, and W. Xiang, ``Contrastive learning for automotive mmWave radar detection points based instance segmentation,'' in \emph{Proceedings of the IEEE International Conference on Intelligent Transportation Systems (ITSC)}, 2022, pp. 1255-1261.


\bibitem{Radar_heatmap_object_detection_and_point_target_tracking} 
P. Li, P. Wang, K. Berntorp, and H. Liu, ``Exploiting temporal relations on radar perception for autonomous driving," in \emph{Proceedings of the IEEE/CVF Conference on Computer Vision and Pattern Recognition (CVPR)}, 2022, pp. 17071-17080.

\bibitem{Radar_LiDAR_fusion_object_detection_RaLiBEV} Y. Yang, J. Liu, T. Huang, Q.-L. Han, G. Ma, and B. Zhu, ``RaLiBEV: Radar and LiDAR BEV fusion learning for anchor box free object detection systems," \textit{arXiv preprint}, 2022. [Online]. Available: \url{arxiv.org/abs/2211.06108}.

\bibitem{Radar_Tracking_ITSC_2022}
T. Zhou et al., ``3D multiple object tracking with multi-modal fusion of low-cost sensors for autonomous driving," in \emph{Proceedings of the IEEE International Conference on Intelligent Transportation Systems (ITSC)}, 2022, pp. 1750-1757.






\bibitem{intro_4d_radar} D. Schwarz, N. Riese, I. Dorsch, and C. Waldschmidt, ``System performance of a 79 GHz high-resolution 4D imaging MIMO radar with 1728 virtual channels,'' \emph{IEEE Journal of Microwaves}, vol. 2, no. 4, pp. 637-647, Oct. 2022.

\bibitem{4D_radar_overview}
Z. Han et al., ``4D millimeter-wave radar in autonomous driving: A survey,'' \textit{arXiv preprint}, 2023. [Online]. Available: \url{arxiv.org/abs/2306.04242}.





\bibitem{model_based_and_DL_based_MOT} J. Pinto, G. Hess, W. Ljungbergh, Y. Xia, H. Wymeersch and L. Svensson, ``Deep learning for model-based multi-object tracking," in \emph{IEEE Transactions on Aerospace and Electronic Systems}, vol. 59, no. 6, pp. 7363-7379, Dec. 2023.

\bibitem{3DMOTFormer}
S. Ding, E. Rehder, L. Schneider, M. Cordts, J. Gall, ``3DMOTFormer: Graph transformer for online 3D multi-object tracking," in \emph{Proceedings of the IEEE/CVF International Conference on Computer Vision (ICCV)}, 2023, pp. 9784-9794.




\bibitem{ACK3DMOT}
G. Guo and S. Zhao, ``3D multi-object tracking with adaptive cubature Kalman filter for autonomous driving," \emph{IEEE Transactions on Intelligent Vehicles}, vol. 8, no. 1, pp. 84-94, Jan. 2023.

\bibitem{SimpleTrack} 
Z. Pang, Z. Li, and N. Wang, ``SimpleTrack: Understanding and rethinking 3D multi-object tracking,'' in \emph{Proceedings of the European Conference on Computer Vision (ECCV) Workshop}, 2022, pp. 680–696.

\bibitem{PF-MOT} 
T. Wen, Y. Zhang, and N. M. Freris, ``PF-MOT: Probability fusion based 3D multi-object tracking for autonomous vehicles," in \emph{Proceedings of the International Conference on Robotics and Automation (ICRA)}, 2022, pp. 700-706.




\bibitem{LiDAR_based_point_target_MOT_PMBM} 
S. Pang, D. Morris, and H. Radha, ``3D multi-object tracking using random finite set-based multiple measurement models filtering (RFS-M3) for autonomous vehicles," in \emph{Proceedings of the 2021 IEEE International Conference on Robotics and Automation (ICRA)}, 2021, pp. 13701-13707.

\bibitem{LiDAR_based_point_target_MOT_GNN-PMB}
J. Liu, L. Bai, Y. Xia, T. Huang, B. Zhu, and Q.-L. Han, ``GNN-PMB: A simple but effective online 3D multi-object tracker without bells and whistles," \emph{IEEE Transactions on Intelligent Vehicles}, vol. 8, no. 2, pp. 1176-1189, Feb. 2023.

\bibitem{LiDAR_based_point_target_MOT_LEGO} Z. Zhang, J. Liu, Y. Xia, T. Huang, Q.-L. Han, and H. Liu, ``LEGO: Learning and graph-optimized modular tracker for online multi-object tracking with point clouds," \textit{arXiv preprint}, 2023. [Online]. Available: \url{arxiv.org/abs/2308.09908}.






\bibitem{Extended_Target_Tracking_Overview}
K. Granstrom, M. Baum, and S. Reuter, ``Extended object tracking: Introduction, overview and applications," \textit{arXiv preprint}, 2016. [Online]. Available: \url{arxiv.org/abs/1604.00970}.

\bibitem{Extended_Target_MOT_GGIW-CPHD}
C. Lundquist, K. Granström, and U. Orguner, ``An extended target CPHD filter and a gamma Gaussian inverse Wishart implementation,'' \emph{IEEE Journal of Selected Topics in Signal Processing}, vol. 7, no. 3, pp. 472-483, Jun. 2013.

\bibitem{Extended_Target_MOT_GGIW-PMBM}
K. Granström, M. Fatemi, and L. Svensson, ``Poisson multi-Bernoulli mixture conjugate prior for multiple extended target filtering,'' \emph{IEEE Transactions on Aerospace and Electronic Systems}, vol. 56, no. 1, pp. 208-225, Feb. 2020.

\bibitem{Extended_Target_MOT_GGIW-PMBM_2}
Á. F. García-Fernández, J. L. Williams, L. Svensson, Y. Xia, ``A Poisson multi-Bernoulli mixture filter for coexisting point and extended Targets,'' \emph{IEEE Transactions on Signal Processing}, vol. 69, pp. 2600-2610, Apr. 2021.

\bibitem{Extended_Target_MOT_GGIW-PMB}
Y. Xia, K. Granström, L. Svensson, M. Fatemi, Á. F. García-Fernández, and J. L. Williams, ``Poisson multi-Bernoulli approximations for multiple extended object filtering'' \emph{IEEE Transactions on Aerospace and Electronic Systems}, vol. 58, no. 2, pp. 890-906, Sep. 2021.

\bibitem{Extended_Target_MOT_BGGIW}
X. Yang, and Q. Jiao, ``Variational approximation for adaptive extended target tracking in clutter with random matrix,'' \emph{IEEE Transactions on Vehicular Technology}, vol. 72, no. 10, pp. 12639-12652, Oct. 2023.

\bibitem{Extended_Target_MOT_EO-GM-PHD}
B. Liu, R. Tharmarasa, R. Jassemi, D. Brown, and T. Kirubarajan, ``RFS-based multiple extended target tracking with resolved multipath detections in clutter,'' \emph{IEEE Transactions on Intelligent Transportation Systems}, vol. 24, no. 10, pp. 10400-10409, Oct. 2023.









\bibitem{LiDAR_Extended_Target_SOT_Rectangle_with_reference_points_GM-PHD_1}
K. Granström, S. Reuter, D. Meissner, and A. Scheel, ``A multiple model PHD approach to tracking of cars under an assumed rectangular shape,'' in \emph{Proceedings of the IEEE 17th International Conference on Information Fusion (FUSION)}, 2014, pp. 1-8.

\bibitem{LiDAR_Extended_Target_SOT_Rectangle_with_reference_points_GM-PHD_2} 
P. Dahal, S. Mentasti, S. Arrigoni, F. Braghin, M. Matteucci, and F. Cheli, ``Extended object tracking in curvilinear road coordinates for autonomous driving," \emph{IEEE Transactions on Intelligent Vehicles}, vol. 8, no. 2, pp. 1266-1278, Feb. 2023.

\bibitem{LiDAR_3D_Extended_Target_SOT_2} 
M. Kumru, and E. Özkan, ``Three-dimensional extended object tracking and shape learning using Gaussian processes," \emph{IEEE Transactions on Aerospace and Electronic Systems}, vol. 57, no. 5, pp. 2795-2814, Oct. 2021.

\bibitem{LiDAR_3D_Extended_Target_SOT_1}
A. Scheel, K. Granstrom, D. Meissner, S. Reuter, and K. Dietmayer, ``Tracking and data segmentation using a GGIW filter with mixture clustering," in \emph{Proceedings of the 17th IEEE International Conference on Information Fusion (FUSION)}, 2014, pp. 1-8.

\bibitem{radar_3D_Extended_Target_SOT}
Y. Xia et al., ``Learning-based extended object tracking using hierarchical truncation measurement model with automotive radar,'' \emph{IEEE Journal of Selected Topics in Signal Processing}, vol. 15, no. 4, pp. 1013-1029, Jun. 2021.

\bibitem{radar_3D_Extended_Target_SOT_2}
X. Cao, J. Lan, X. R. Li, and Y. Liu, ``Automotive radar-based vehicle tracking using data-region association,'' \emph{IEEE Transactions on Intelligent Transportation Systems}, vol. 23, no. 7, pp. 8997-9010, July. 2022.

\bibitem{PMRA-PMBM} G. Ding, J. Liu, Y. Xia, T. Huang, B. Zhu, and J. Sun, ``LiDAR point cloud-based multiple vehicle tracking with probabilistic measurement-region association," \textit{arXiv preprint}, 2024. [Online]. Available: \url{arxiv.org/abs/2403.06423}.

\bibitem{LiDAR_3D_Extended_Target_MOT_1}
K. Granström, L. Svensson, S. Reuter, Y. Xia, and M. Fatemi, ``Likelihood-based data association for extended object tracking using sampling methods," \emph{IEEE Transactions on Intelligent Vehicles}, vol. 3, no. 1, pp. 30-45, Mar. 2018.

\bibitem{LiDAR_3D_Extended_Target_MOT_2}
F. Meyer, and J. L. Williams, ``Scalable detection and tracking of geometric extended objects,'' \emph{IEEE Transactions on Signal Processing}, vol. 69, no. 1, pp. 6283-6298, Oct. 2021.


\bibitem{SimTrack} 
C. Luo, X. Yang, and A. Yuille, ``Exploring simple 3D multi-object tracking for autonomous driving," in \emph{Proceedings of the IEEE/CVF International Conference on Computer Vision (ICCV)}, 2021, pp. 10488-10497.

\bibitem{CenterTube} 
H. Liu, Y. Ma, Q. Hu, and Y. Guo, ``CenterTube: Tracking multiple 3D objects with 4D tubelets in dynamic point clouds,'' \emph{IEEE Transactions on Multimedia}, vol. 25, pp. 8793-8804, Feb. 2023.







\bibitem{3DMODT}
J. Kini, A. Mian, and M. Shah, ``3DMODT: Attention-guided affinities for joint detection and tracking in 3D point clouds," in \emph{Proceedings of the IEEE International Conference on Robotics and Automation (ICRA)}, 2023, pp. 841-848.

\bibitem{MUTR3D}
T. Zhang, X. Chen, Y. Wang, Y. Wang, and H. Zhao, ``{MUTR3D}: A multi-camera tracking framework via 3D-to-2D queries,’’ in \emph{Proceedings of the IEEE/CVF Conference on Computer Vision and Pattern Recognition (CVPR) Workshop}, 2022, pp. 4537-4546.

\bibitem{Trackformer}
T. Meinhardt, A. Kirillov, L. Leal-Taixe, and C. Feichtenhofer, ``Trackformer: Multi-object tracking with transformers,’’ in \emph{Proceedings of the IEEE/CVF Conference on Computer Vision and Pattern Recognition (CVPR)}, 2022, pp. 8844-8854.


\bibitem{4d_imaging_radar_3d_object_detection_2} B. Xu et al., ``RPFA-Net: a 4D RaDAR pillar feature attention network for 3D object detection,'' in \emph{Proceedings of the IEEE International Conference on Intelligent Transportation Systems (ITSC)}, 2021, pp. 3061-3066.

\bibitem{4d_imaging_radar_3d_object_detection_3} B. Tan et al., ``3D object detection for multi-frame 4D automotive millimeter-wave radar point cloud,'' \emph{IEEE Sensors Journal}, vol. 23, no. 11, pp. 11125-11138, Jun. 2023.

\bibitem{SMURF} J. Liu, Q. Zhao, W. Xiong, T. Huang, Q.-L. Han, and B. Zhu, ``SMURF: Spatial multi-representation fusion for 3D object detection with 4D imaging radar," \emph{IEEE Transactions on Intelligent Vehicles}, vol. 9, no. 1, pp. 799-812, Jan. 2024.

\bibitem{VoD_dataset} A. Palffy, E. Pool, S. Baratam, J. F. P. Kooij, and D. M. Gavrila, ``Multi-class road user detection with 3+1D radar in the View-of-Delft dataset,'' \emph{IEEE Robotics and Automation Letters}, vol. 7, no. 2, pp. 4961-4968, Apr. 2022.

\bibitem{TJ4DRadSet_dataset} L. Zheng et al., ``TJ4DRadSet: A 4D radar dataset for autonomous driving,'' in \emph{Proceedings of the IEEE International Conference on Intelligent Transportation Systems (ITSC)}, 2022, pp. 493-498.


\bibitem{4d_imaging_radar_camera_3d_object_detection_rcfusion} L. Zheng et al., ``RCFusion: Fusing 4D radar and camera with bird’s-eye view features for 3D object detection,'' \emph{IEEE Transactions on Instrumentation and Measurement}, vol. 72, pp. 1-14, May 2023.


\bibitem{4d_imaging_radar_camera_3d_object_detection_LXL} W. Xiong, J. Liu, T. Huang, Q.-L. Han, Y. Xia, and B. Zhu, ``LXL: LiDAR excluded lean 3D object detection with 4D imaging radar and camera fusion,'' \emph{IEEE Transactions on Intelligent Vehicles}, vol. 9, no. 1, pp. 79-92, Jan. 2024.

\bibitem{4d_imaging_radar_lidar_3d_object_detection_1} L. Wang et al., ``InterFusion: Interaction-based 4D radar and LiDAR fusion for 3D object detection,'' in \emph{Proceedings of the IEEE/RSJ International Conference on Intelligent Robots and Systems (IROS)}, 2022, pp. 12247-12253.

\bibitem{4d_imaging_radar_lidar_3d_object_detection_2} L. Wang et al., ``Multi-modal and multi-scale fusion 3D object detection of 4D radar and LiDAR for autonomous driving,'' \emph{IEEE Transactions on Vehicular Technology}, vol. 72, no. 5, pp. 5628-5641, May 2023.

\bibitem{Point_target_MOT_BP} F. Meyer et al., ``Message passing algorithms for scalable multitarget tracking," \emph{Proceedings of the IEEE}, vol. 106, no. 2, pp. 221–259, Feb. 2018.







\bibitem{CAMO-MOT}
L. Wang et al., ``CAMO-MOT: Combined appearance-motion optimization for 3D multi-object tracking with camera-LiDAR fusion,'' \emph{IEEE Transactions on Intelligent Transportation Systems}, vol. 24, no. 11, pp. 11981-11996, Nov. 2023.


\bibitem{C_R_Track_Level_Fusion}
X. Hao, Y. Xia, H. Yang, and Z. Zuo, ``Asynchronous information fusion in intelligent driving systems for target tracking using cameras and radars,'' \emph{IEEE Transactions on Industrial Electronics}, vol. 70, no. 3, pp. 2708-2717, Mar. 2023.












\bibitem{global_hypothesis_definition}
J. L. Williams, ``Marginal multi-Bernoulli filters: RFS derivation of MHT, JIPDA, and association-based MeMBer,'' \emph{IEEE Transactions on Aerospace and Electronic Systems}, vol. 51, no. 3, pp. 1664-1687, Jul. 2015.





\bibitem{DBSCAN}
E. Schubert, J. Sander, M. Ester, H. P. Kriegel, and X. Xu, ``DBSCAN revisited, revisited: why and how you should (still) use DBSCAN,'' \emph{ACM Transactions on Database Systems (TODS)}, vol. 42, no. 3, pp. 1-21, Sep. 2017.

\bibitem{k-means}
M. Ahmed, R. Seraj, and S. M. S. Islam, ``The k-means algorithm: A comprehensive survey and performance evaluation,'' \emph{Electronics}, vol. 9, no. 8, pp. 1295, Aug. 2020.



\bibitem{point_target_MOT_PMBM} 
Á. F. García-Fernández, J. L. Williams, K. Granström, and L. Svensson, ``Poisson multi-Bernoulli mixture filter: Direct derivation and implementation,'' \emph{IEEE Transactions on Aerospace and Electronic Systems}, vol. 54, no. 4, pp. 1883-1901, Aug. 2018.

\bibitem{NMS} A. Neubeck and L. Van Gool, ``Efficient non-maximum suppression," in \emph{Proceedings of the International Conference on Pattern Recognition (ICPR)}, 2006, vol. 3, pp. 850-855.























\bibitem{HOTA} J. Luiten et al., ``HOTA: A higher order metric for evaluating multi-object tracking," \emph{International Journal of Computer Vision}, vol. 129, pp. 548-578, Oct. 2020.

\bibitem{nuScenes} H. Caesar et al., ``nuScenes: A multimodal dataset for autonomous driving," in \emph{Proceedings of the IEEE/CVF Conference on Computer Vision and Pattern Recognition (CVPR)}, 2020, pp. 11621-11631.










\end{thebibliography}
\end{document}